\documentclass[a4paper]{article}

\usepackage[braket, qm]{qcircuit}
\usepackage[utf8]{inputenc}
\usepackage{hyperref}
\usepackage{amsmath,amssymb,amsthm}
\usepackage{braket}
\usepackage{tikz}
\usepackage{graphicx}
\usepackage{subcaption}
\usepackage{mathdots}

\newtheorem{theorem}{Theorem}[section]

\newtheorem{definition}[theorem]{Definition}
\newtheorem{remark}[theorem]{Remark}
\newtheorem{example}[theorem]{Example}
\newtheorem*{note}{Note}

\numberwithin{equation}{section}

\title{Quantum simulation of perfect state transfer on weighted cubelike graphs}
\author{Jaideep Mulherkar \\ jaideep\_mulherkar@daiict.ac.in\and Rishikant Rajdeepak \\201521006@daiict.ac.in \and V. Sunita \\ v\_suni@daiict.ac.in}

\begin{document}
\maketitle

\begin{abstract}
    A continuous-time quantum walk on a graph evolves according to the unitary operator $e^{-iAt}$, where $A$ is the adjacency matrix of the graph. Perfect state transfer (PST) in a quantum walk is the transfer of a quantum state from one node of a graph to another node with $100\%$ fidelity. It can be shown that the adjacency matrix of a cubelike graph is a finite sum of tensor products of Pauli $X$ operators. We use this fact to construct an efficient quantum circuit for the quantum walk on cubelike graphs. In \cite{Cao2021,rishi2021(2)}, a characterization of integer weighted cubelike graphs is given that exhibit periodicity or PST at time $t=\pi/2$. We use our circuits to demonstrate PST or periodicity in these graphs on IBM's quantum computing platform~\cite{Qiskit,IBM2021}.
\end{abstract}  

\textbf{keywords:} Continuous-time quantum walk, Perfect state transfer, Periodicity, Quantum circuits.

\section{Introduction}
A quantum random walk is the quantum analogue of a classical random walk \cite{Kempe2003,renatobook,andraca2012}. The study of classical random walks has led to many applications in science and engineering, such as in the study of randomized algorithms and a sampling approach called Markov chain Monte Carlo in computer science, in the study of social networks, in the behavior of stock prices in finance, in models of diffusion and study of polymers in Physics, and the motion of motile bacteria in biology.  In \cite{aharonov1993,farhi1998}, the first models for quantum random walks were proposed. Since then, quantum walks have been a source of intense study. Researchers observed that there are some startling differences between classical and quantum walks. For example, a quantum walk on a one-dimensional lattice spreads quadratically faster than a classical walk  \cite{nayak2000}. Quantum walks on cubelike graphs, such as the hypercubes, hit exponentially faster to the antipodal vertex as compared to classical counterparts \cite{Kempe2005}. 

Quantum walks on graphs are of two types: discrete and continuous. In the discrete case, a graph is associated with a Hilbert space of dimension $N\times \Delta$, where $N$ is the number of vertices, and $\Delta$ is the maximum degree of the graph. In the continuous case, a graph is associated with a Hilbert space of dimension $N$, and the evolution of the system is described by $e^{\iota tA}$, where $A$ is the adjacency matrix of the graph and $t$ is real. An essential feature of a quantum walk is a quantum state transfer from one vertex to another with high fidelity. When the transfer occurs with $100 \%$ fidelity, it is called perfect state transfer (PST). Some of the excellent survey papers on graph families that admit PST are ~\cite{godsil2012,godsil2011}. Among these graphs, cubelike graphs are the most famous ones that have been researched thoroughly for determining the existence and finding the pair of vertices admitting perfect state transfer in constant time \cite{bernasconi2008,cheung2011}. Notice that all cubelike graphs do not allow perfect state transfer. The study of PST on weighted graphs has been less studied. Recently, weighted abelian Cayley graphs have been characterized that exhibit PST ~\cite{Cao2021}.

In this paper, we look at the implementation of Perfect state transfer on weighed cubelike graphs. Some of the efficient implementation of quantum walks are described in \cite{frank2020,abhijith2020,rishi2021,wanzambi2021,warat2021}. It can be shown that the adjacency matrix of a cubelike graph is the sum of the tensor products of Pauli $X$ operators. One then observes that generating efficient quantum circuits for quantum walks can then be done by quantum hamiltonian simulation techniques that have been described in \cite{nielsen2011}. We use quantum simulation techniques to verify the theoretical results of PST on weighted cubelike graphs.

\section{Preliminaries}
An undirected weighted graph $\Gamma$ consists of a triplet $(V,E,f)$, where $V$ is a non-empty set whose elements are called vertices, $E$ is a set of edges, where an edge is an unordered tuple of vertices, and $f:V\times V\rightarrow \mathbb{R}$ is a weight function that assigns non-zero real weights to edges. If $\Gamma$ is finite, then its adjacency matrix $A$ is defined by; 
\begin{equation*}
    A_{u,v} = f((u,v)), \qquad (u,v)\in V\times V.
\end{equation*}
The adjacency matrix $A$ is real and symmetric. A tuple $(u,u)$ is a loop if its weight is non-zero. If $f((u,u))=0$ for all $u\in V$, then the diagonal entries of $A$ are zero and the graph has no loops. A graph family of interest is a weighted cubelike graph which is defined as follows.

\begin{definition}
Let $f$ be a real-valued function over a finite Boolean group $\mathbb{Z}_2^n$ of dimension $n>0$. A cubelike graph, denoted by $Cay(\mathbb{Z}_2^n,f)$, is a graph with vertex-set $\mathbb{Z}_2^n$, and the weight of a pair $(u,v)$ of vertices is given by $f(u\oplus v)$, where $\oplus$ denotes the group addition, i.e., componentwise addition modulo 2. The adjacency matrix $A$ of $Cay(\mathbb{Z}_2^n,f)$ is given by; \[A_{u,v}=f(u\oplus v),\; u,v\in V.\]
\end{definition}
An equivalent definition for an unweighted cubelike graph is given by; let $\Omega_f=\{u\in\mathbb{Z}_2^n: f(u)=1\}$, then two vertices $u$ and $v$ are adjacent if $u\oplus v\in \Omega_f$. The cubelike graph, in this case, is denoted by $Cay(\mathbb{Z}_2^n,\Omega_f)$, see Fig.~\ref{fig:Q3} and Fig.~\ref{fig:AQ3}. 
\begin{figure}
\tiny
\centering
    \begin{minipage}[t]{.45\textwidth}
    \centering
    \begin{tikzpicture}[scale=.6]
        \begin{scope}[every node/.style = {draw, fill, circle, inner sep=1pt}]
        \node (0) at (-1,1) [label=left:$000$]{};
        \node (1) at (1,1) [label=right:$001$]{};
        \node (3) at (1,-1) [label=right:$011$]{};
        \node (2) at (-1,-1) [label=left:$010$]{};
        \node (4) at (-3,3) [label=left:$100$]{};
        \node (5) at (3,3) [label=right:$101$]{};
        \node (7) at (3,-3) [label=right:$111$]{};
        \node (6) at (-3,-3) [label=left:$110$]{};
    \end{scope}
        \draw (0) edge  (1); 
        \draw (0) edge  (2); 
        \draw (0) edge  (4); 
        \draw (1) edge  (3); 
        \draw (1) edge  (5); 
        \draw (2) edge  (3);
        \draw (2) edge  (6); 
        \draw (3) edge  (7); 
        \draw (4) edge  (5); 
        \draw (4) edge  (6); 
        \draw (5) edge  (7); 
        \draw (6) edge  (7); 
    \end{tikzpicture}
    \caption{\label{fig:Q3} $Cay(\mathbb{Z}_2^3,\{001,010,100\})$}
    \end{minipage}
    \begin{minipage}[t]{.45\textwidth}
    \centering
    \begin{tikzpicture}[scale=.6]
        \begin{scope}[every node/.style = {draw, circle, inner sep=1pt}]
        \node (0) at (-1,1) [label=left:$000$]{};
        \node (1) at (1,1) [label=right:$001$]{};
        \node (3) at (1,-1) [label=right:$011$]{};
        \node (2) at (-1,-1) [label=left:$010$]{};
        \node (4) at (-3,3) [label=left:$100$]{};
        \node (5) at (3,3) [label=right:$101$]{};
        \node (7) at (3,-3) [label=right:$111$]{};
        \node (6) at (-3,-3) [label=left:$110$]{};
    \end{scope}
        \draw (0) edge (1); 
        \draw (0) edge (2); 
        \draw (0) edge  (4);
        \draw (0) edge[bend left=20] (3);
        \draw (0) edge[bend right=20] (7);
        \draw (1) edge  (3); 
        \draw (1) edge (5);
        \draw (1) edge[bend right=20]  (2);
        \draw (1) edge [bend left=20]  (6);
        \draw (2) edge  (3);
        \draw (2) edge  (6);
        \draw (2) edge[bend right=25] (5);
        \draw (3) edge[bend left=25] (4);
        \draw (3) edge   (7); 
        \draw (4) edge   (5);
        \draw (4) edge  (6);
        \draw (4) edge[bend right=55] (7);
        \draw (5) edge  (7); 
        \draw (5) edge[ bend left=55]  (6);
        \draw (6) edge  (7); 
    \end{tikzpicture}
    \caption{\label{fig:AQ3} $Cay(\mathbb{Z}_2^3,\{001,010,011,100,111\})$}
    \end{minipage}
\end{figure}

\subsection{Continuous-time quantum walk}
Let $\Gamma$ be an undirected and weighted graph with or without loops and $A$ be the adjacency matrix. A quantum walk on $\Gamma$ is described by an evolution of the quantum system associated with the graph. Suppose the graph has $N$ vertices, then it is associated with a Hilbert space $\mathcal{H}_P\cong \mathbb{C}^N$ of dimension $N$, called the position space, and the computational basis is represented by;
\[
\{\ket{v}:v\mbox{ is a vertex in }\Gamma\}.
\]
The continuous-time quantum walk (CTQW) on $\Gamma$ is described by the transition matrix $\mathcal{U}(t)=e^{-\iota t A}$, where $\iota=\sqrt{-1}$, that operates on the position space $\mathcal{H}_P$. In other words, if $\ket{\psi(0)}$ is the initial state of the quantum system associated with the graph, then the state of the system after time $t$ is given by 
\[
\ket{\psi(t)}=e^{-\iota t A}\ket{\psi(0)}.
\]
\begin{definition}
A graph is said to admit perfect state transfer (PST) if the quantum walker beginning at some vertex $u$ reaches a distinct vertex $v$ with probability $1$, i.e., for some positive real $\tau$ and scalar $\lambda$
\[
|\braket{v|e^{-\iota \tau A}|u}|=|\lambda|=1.
\]
Alternatively, we say perfect state transfer occurs from the vertex $u$ to the vertex $v$. If $u=v$, we say the graph is periodic at $u$ with period $\tau$, and if the graph is periodic at every vertex with the same period $\tau$ then, the graph is periodic.
\end{definition}

\begin{example}\label{eg:cycle}
Consider the graph on the cycle of size 4, see Fig.~\ref{fig:cycle}, with the adjacency matrix $A$ given by;
\[
A=\begin{bmatrix} 0 & 1 & 1 & 0 \\ 1 & 0 & 0 & 1 \\ 1&0&0&1 \\ 0&1&1&0  \end{bmatrix}.
\]
Then, the transition matrix at time $t=\pi/2$ is
\[
\mathcal{U}(t=\frac{\pi}{2}) = \begin{bmatrix} 0&0&0&1 \\ 0&0&1&0 \\0&1&0&0 \\ 1&0&0&0 \end{bmatrix}.
\]
Thus, perfect state transfer occurs between the pairs $\{1,4\}$ and $\{2,3\}$, both in time $\frac{\pi}{2}$. The graph is periodic with period $\pi$.
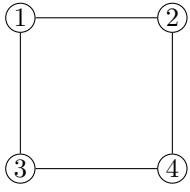
\begin{figure}
    \centering
    \begin{tikzpicture}
        \begin{scope}[every node/.style={draw,circle,inner sep=1pt}]
            \node (0) at (-1,1) {1};
            \node (1) at (1,1) {2};
            \node (3) at (1,-1) {4};
            \node (2) at (-1,-1) {3};
        \end{scope} 
        \draw (0) -- (1); \draw (0) -- (2); \draw (1) -- (3); \draw (2) -- (3);
    \end{tikzpicture}
    \caption{PST occurs between the pairs \{1,4\} and \{2,3\} with time $\frac{\pi}{2}$, and the graph is periodic with period $\pi$.}\label{fig:cycle}
\end{figure}
\end{example}   

\subsection{Decomposition of the adjacency matrix of weighted cubelike graph}
\subsubsection{Group representations}
An $m$-degree representation of a finite group $G$ is a homomorphism $\rho$ from $G$ into the general linear group $GL(V)$ of an $m$-dimensional vector space $V$ over the field $\mathbb{F}$, where $\mathbb{F}$ is a complex or real field. Since $GL(V)$ is isomorphic to $GL_m(\mathbb{F})$, the general linear group of degree $m$ that consists of $m\times m$ invertible matrices, an equivalent definition for the group representation is the group homomorphism 
\[
\rho:G\rightarrow GL_m(\mathbb{F}).
\] 
The group algebra $\mathbb{C}[G]$ is an inner product space whose vectors are formal linear combinations of the group elements, i.e.,
\[
\mathbb{C}[G]=\left\{\sum_{g\in G}\lambda_g g : \lambda_g\in\mathbb{C}\right\},
\]
with the vector addition, the scalar multiplication, and the inner product defined by;
\[
\begin{split}
    \sum_{g\in G}\lambda_g g + \sum_{g\in G}\mu_g g &= \sum_{g\in G}(\lambda_g+\mu_g)g, \qquad\mbox{(addition)}, \\
    \lambda \sum_{g\in G}\lambda_g g &= \sum_{g\in G}(\lambda\lambda_g)g \qquad\mbox{(scalar multiplication)},\\
    \left\langle \sum_{g\in G}\lambda_gg,\sum_{g\in G}\mu_gg \right\rangle &= \sum_{g\in G}\lambda_g\bar{\mu}_g, \qquad \mbox{(inner product)}.
\end{split} 
\]
The regular representation on $G$, $\rho_{reg}:G\rightarrow GL(\mathbb{C}[G])$, is defined by;
\[
\rho_{reg} (x) \left(\sum_{g\in G}\lambda_gg\right) = \sum_{g\in G}\lambda_g (xg) = \sum_{y\in G} \lambda_{x^{-1}y}y.
\]

\subsubsection{The decomposition}
If $G=\mathbb{Z}_2^n$, then for $x\in\mathbb{Z}_2^n$ the regular representation acts on $\mathbb{Z}_2^n$ as
\begin{equation*}
\rho_{reg}(x)y = x\oplus y = (x_1\oplus y_1, \dots, x_n\oplus y_n), \qquad y\in\mathbb{Z}_2^n.
\end{equation*}
Let $X$, $Y$ and $Z$ denote the three Pauli matrices that acts on the computational basis $\{\ket{0},\ket{1}\}$ of the two dimensional Hilbert space $\mathbb{C}^2$ as
\begin{equation*}
    X\ket{a} = \ket{a\oplus 1},\; Y\ket{a}=(-1)^a\iota \ket{a\oplus 1}, \; Z\ket{a} = (-1)^a \ket{a}, \; a\in \{0,1\}.
\end{equation*}
The group element $y$ is also a vector in $\mathbb{C}[\mathbb{Z}_2^n]$ whose matrix representation is $\ket{y}=\ket{y_1}\otimes \cdots \otimes \ket{y_n}$. Hence, the action of $\rho_{reg}(x)$ over $y$ can be rewritten as 
\[
\begin{split}
\rho_{reg}(x)y &= (X^{x_1}\ket{y_1})\otimes \cdots \otimes (X^{x_n}y_n), \mbox{ where }X^{x_i}\ket{y_i}=\ket{x_i\oplus y_i}, \\
&= \left(X^{x_1}\otimes \cdots \otimes X^{x_n}\right)(\ket{y_1}\otimes\cdots\otimes \ket{y_n}).
\end{split}
\]
The adjacency matrix $A$ of $Cay(\mathbb{Z}_2^n,f)$ is decomposed by using the regular representation on $\mathbb{Z}_2^n$, viz., given $x,y\in\mathbb{Z}_2^n$, the value $\rho_{reg}(x)y=x\oplus y$ corresponds to the $(x,y)$-entry of $A$, so $A$ can be expressed as;
\begin{equation}
    A = \sum_{x\in\mathbb{Z}_2^n}f(x)\rho_{reg}(x).
\end{equation}
Since $\rho_{reg}(x)$ commutes with $\rho_{reg}(y)$ for all $x,y\in\mathbb{Z}_2^n$, the evolution operator $\mathcal{U}(t)=e^{-\iota t A}$ is decomposed into;
\begin{equation}\label{eq:evolution}
    \mathcal{U}(t) = \prod_{x\in\mathbb{Z}_2^n} U(x,t), \qquad U(x,t)=e^{-\iota t f(x)\rho_{reg}(x)}.
\end{equation}

\subsection{PST or periodicity in weighted cubelike graphs}
We simulate continuous-time quantum walk on $Cay(\mathbb{Z}_2^n,f)$ and verify the existence of perfect state transfer or periodicity as mentioned in the following theorem.
\begin{theorem}\cite{Cao2021,rishi2021(2)}\label{thm:pst}
Let $f:\mathbb{Z}_2^n\rightarrow \mathbb{Z}$ be an integer-valued function. For $x\in\mathbb{Z}_2^n$, define a subset $O_x=\{y\in\mathbb{Z}_2^n: \braket{x|y}\mbox{ mod }2=1\}$. Let $e_i$, $1\leq i\leq n$, denote the $n$-tuple with entry 1 at position $i$ and zero everywhere else. Let $\sigma\in\mathbb{Z}_2^n$ such that
\begin{equation}\label{eq:sigma}
\sigma_i = 1 \mbox{ only if } \sum_{y\in O_{e_i}} f(y)\mbox{ mod }2=1.
\end{equation}
Then, 
\begin{enumerate}
    \item if $\sigma$ is the identity element, i.e., $\sigma=\textbf{0}$, then $Cay(\mathbb{Z}_2^n,f)$ is periodic with period $\frac{\pi}{2}$,
    \item if $\sigma\neq \textbf{0}$, then PST occurs between every pair $\{u,v\}$ satisfying $u\oplus v=\sigma$, with time $\tau=\frac{\pi}{2}$.
\end{enumerate} 
\end{theorem}   

\begin{note}
Although PST or periodicity in weighted cubelike graph mentioned in \cite{rishi2021(2)} was done independently, it was only later that the authors realized that its generalized version, viz., PST on weighted abelian Cayley graph, has already been proved in another paper \cite{Cao2021}. 
\end{note}

\section{The quantum simulation}
\begin{figure}
\begin{equation*}
    \Qcircuit @C=1em @R=1.5em
    {
    \lstick{} & \qw & \ctrl{3} & \qw  & \qw & \qw & \qw & \qw & \ctrl{3} & \qw & \qw \\
    \lstick{} & \qw & \qw & \ctrl{2}  & \qw & \qw & \qw & \ctrl{2} & \qw & \qw & \qw \\
    \lstick{} & \qw & \qw & \qw & \ctrl{1} & \qw & \ctrl{1} & \qw & \qw & \qw & \qw \\
    \lstick{\ket{0}} & \qw & \targ & \targ & \targ & \gate{e^{-\iota t Z}} & \targ & \targ & \targ & \qw & \qw & \rstick{\ket{0}}
    }
\end{equation*}
\caption{Quantum circuit to implement $e^{-\iota t A}$, where $A=Z\otimes Z\otimes Z$.}
\label{fig:H}
\end{figure}
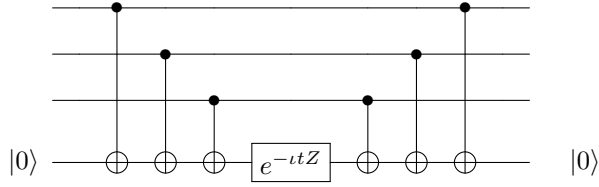

The idea to design a quantum circuit for CTQW on a cubelike graph has been taken from \cite{nielsen2011}; if the Hamiltonian is given by $A=Z_1\otimes \cdots \otimes Z_n$, where $Z_i=Z$, then the phase shift applied to the system is $e^{-\iota t}$ if the parity of the $n$ qubits in the computational basis is even, otherwise, the phase shift applied is $e^{\iota t}$. Fig.~\ref{fig:H} illustrates the quantum circuit for $e^{-\iota t A}$, where $A=Z\otimes Z\otimes Z$.

\subsection{Quantum circuits}
\begin{figure}
\begin{equation*}
   \Qcircuit @C=1em @R=1em
    {
\lstick{\ket{x_1}} & \gate{\mathrm{H}} & \qw & \cdots && \qw & \qw & \qw & \cdots & & \qw & \gate{\mathrm{H}} & \qw \\
\lstick{\vdots} & \vdots &  & \ddots && \vdots & \vdots & \vdots & \iddots & & \vdots & \vdots & \vdots \\
\lstick{\ket{x_{i_1}}} & \gate{\mathrm{H}} & \ctrl{5} & \cdots && \qw & \qw & \qw & \cdots & & \ctrl{5} & \gate{\mathrm{H}} & \qw \\
\lstick{\vdots} & \vdots & & \ddots && \vdots & \vdots & \vdots & \iddots & & & \vdots & \vdots \\
\lstick{\ket{x_{i_k}}} & \gate{\mathrm{H}} & \qw & \cdots && \ctrl{3} & \qw & \ctrl{3} & \cdots & & \qw & \gate{\mathrm{H}} & \qw \\
\lstick{\vdots} & \vdots &  & \ddots & &  & \vdots &  & \iddots & & & \vdots & \vdots \\
\lstick{\ket{x_{n}}} & \gate{\mathrm{H}} & \qw  & \cdots & & \qw & \qw & \qw & \cdots & & \qw & \gate{\mathrm{H}} & \qw \\
\lstick{\ket{0}} & \qw & \targ & \qw & \qw & \targ & \gate{\mathrm{R_{\hat{z}}(2tf(x))}} & \targ & \qw & \qw & \targ & \qw & \qw  
    }
\end{equation*}
    \caption{Quantum circuit for $U(x,t)=e^{-\iota t f(x) \rho_{reg}(x)}$.}
    \label{fig:simulation}
\end{figure}
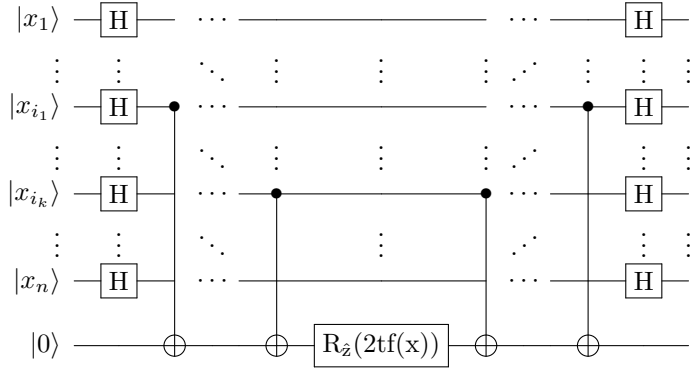

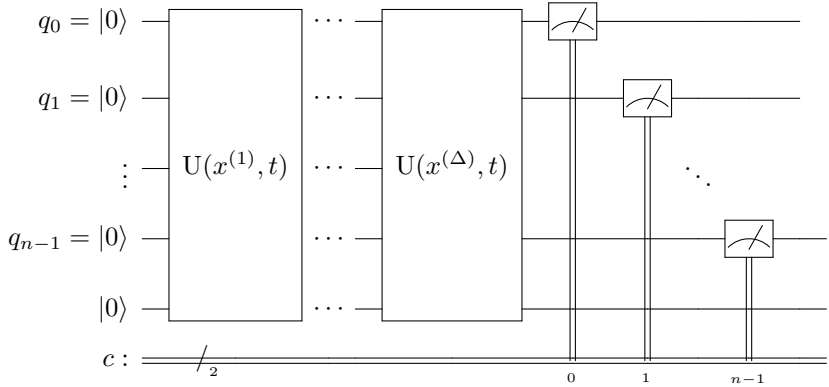
\begin{figure}[t]
\centering
\begin{equation*}
    \Qcircuit @C=1em @R=1.5em
    {
\lstick{q_0=\ket{0}} & \multigate{4}{\mathrm{U}(x^{(1)},t)} & \cdots & & \multigate{4}{\mathrm{U}(x^{(\Delta)},t)} & \meter & \qw & \qw & \qw & \qw \\
\lstick{q_1=\ket{0}} & \ghost{{\mathrm{U}(x^{(1)},t)}} & \cdots & & \ghost{\mathrm{U}(x^{(\Delta)},t)} & \qw &  \meter & \qw & \qw & \qw \\
\lstick{\vdots} & \ghost{{\mathrm{U}(x^{(1)},t)}} & \cdots & & \ghost{\mathrm{U}(x^{(\Delta)},t)} &  &  & \ddots &   &  \\
\lstick{q_{n-1}=\ket{0}} & \ghost{{\mathrm{U}(x^{(1)},t)}} & \cdots & & \ghost{\mathrm{U}(x^{(\Delta)},t)} & \qw & \qw & \qw & \meter & \qw  & \qw \\
\lstick{\ket{0}} & \ghost{{\mathrm{U}(x^{(1)},t)}} & \cdots & & \ghost{\mathrm{U}(x^{(\Delta)},t)} & \qw & \qw & \qw & \qw & \qw & \qw \\
\lstick{c:} & \lstick{/_{_{_2}}} \cw & \cw & \cw & \cw & \dstick{_{_0}} \cwx[-5] \cw & \dstick{_{_1}} \cw \cwx[-4]  & \cw & \dstick{_{_{n-1}}} \cw \cwx[-2] & \cw & \cw 
    }
\end{equation*}
\caption{An illustration of CTQW quantum circuit on weighted cubelike graph}
\label{fig:gen_circuit}
\end{figure}

Let $x\in\mathbb{Z}_2^n$, then the regular representation $\rho_{reg}(x)$ is given by
\begin{equation*}
    \rho_{reg}(x) = \otimes_{i=1}^nX^{x_i} = H^{\otimes n} \left( \otimes_{i=1}^nZ^{x_i} \right) H^{\otimes n}, \mbox{ since } X=HZH.
\end{equation*}
Applying the changes to the operator $U(x,t)$ in Eq.~\ref{eq:evolution}, we get
\begin{equation*}
    \begin{split}
        U(x,t) &= e^{-\iota t f(x) \rho_{reg}(x)} = e^{-\iota t f(x) \left[\otimes_{i=1}^nX^{x_i}\right]} \\
        &= \sum_{l=0}^\infty \frac{(-\iota t f(x))^l}{l!} \left[\otimes_{i=1}^nX^{x_i}\right]^l \\
        &= \sum_{l=0}^\infty \frac{(-\iota t f(x)))^{2l}}{(2l)!} I^{\otimes n} + \sum_{l=0}^\infty \frac{(-\iota t f(x))^{2l+1}}{(2l+1)!} \left[\otimes_{i=1}^nX^{x_i}\right] \\ 
        &= H^{\otimes n}V(x,t)H^{\otimes n}, \qquad V(x,t)=e^{-\iota t f(x) \left[\otimes_{i=1}^n Z^{x_i}\right]}.
    \end{split}
\end{equation*}
We see that, 
\begin{equation*}
\begin{split}
\left(Z_1^{x_1}\otimes \cdots \otimes Z_n^{x_n}\right) \ket{y} &= (-1)^{x_1y_1}\ket{y_1}\otimes \cdots \otimes (-1)^{x_ny_n}\ket{y_n} \\
&= (-1)^{\sum_{i=1}^nx_iy_i}\ket{y_1}\otimes \cdots \otimes \ket{y_n} \\
&= \begin{cases} \ket{y}, & \mbox{if }\braket{x|y}\mbox{ mod }2=0 \\ -\ket{y}, & \mbox{if }\braket{x|y}\mbox{ mod }2=1. \end{cases} 
\end{split}
\end{equation*}
This implies,
\begin{equation*}
    V(x,t)\ket{y} = \begin{cases} e^{-\iota t f(x) Z}\ket{y} & \mbox{ if }\braket{x|y}\mbox{ mod }2=0 \\ e^{\iota t f(x) Z}\ket{y} & \mbox{if }\braket{x|y}\mbox{ mod }2=1. \end{cases}
\end{equation*}
Thus, the action of the operator $V(x,t)$ is equivalent to the application of the rotation operator $R_{\hat{z}}(2tf(x))$ about the $\hat{z}$-axis if $\braket{x|y}$ is even, and $R_{\hat{z}}(-2tf(x))$ if $\braket{x|y}$ is odd. Hence, if $x$ has non-zero entries at positions $i_1,\dots,i_k$, then the quantum circuit for the operator $e^{-\iota tf(x)\rho_{reg}(x)}$ is depicted by Fig.~\ref{fig:simulation}. Suppose elements in $\Omega_f=\{y:f(y)\neq 0\}$ are represented by $\Omega_f=\{x^{(1)},\dots,x^{(\Delta)}\}$, where $\Delta$ is the cardinality of $\Omega_f$, then the quantum circuit for the continuous-time quantum walk is as shown in Fig.~\ref{fig:gen_circuit}, where the initialized state, in general, is $\ket{0}^{\otimes n}$ along with an ancilla qubit with state $\ket{0}$. 

\begin{remark}
    As seen in Fig.~\ref{fig:gen_circuit}, the Hadamard gates $H$ applied at the end of $U(x^{(i)},t)$ and the beginning of $U(x^{(i+1)},t)$, $1\leq i <\Delta$, are not required, because $H^2=I$, thus the actual number of $H$ gates required are $2n$. Secondly, the number of rotation operators used are $\Delta$. Lastly, for each $x\in\Omega_f$, the number of CNOT gates applied are equal to the Hamming weight $wt(x)$ of $x$. Thus, the total number of CNOT gates used are $\sum_{x\in\Omega_f}wt(x)$. 
\end{remark}

\subsection{Results}
\begin{figure}[t]
    \centering
    \scalebox{.6}{\includegraphics[width=20cm]{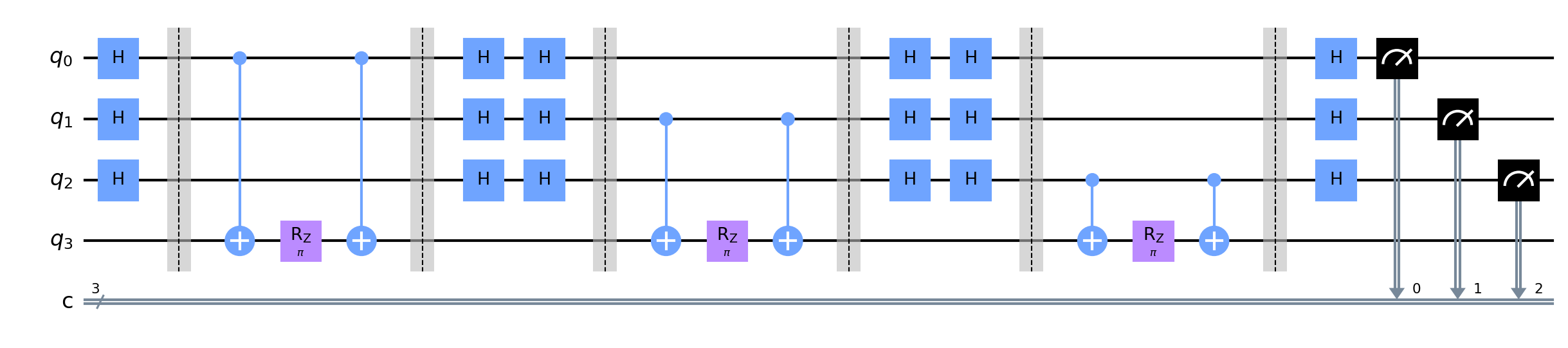}}
    \caption{Quantum circuit for $Cay(\mathbb{Z}_2^3,\{001,010,100\})$.}
    \label{fig:fig7}
\end{figure}

\begin{figure}[t]
    \centering
    \scalebox{.6}{\includegraphics[width=20cm]{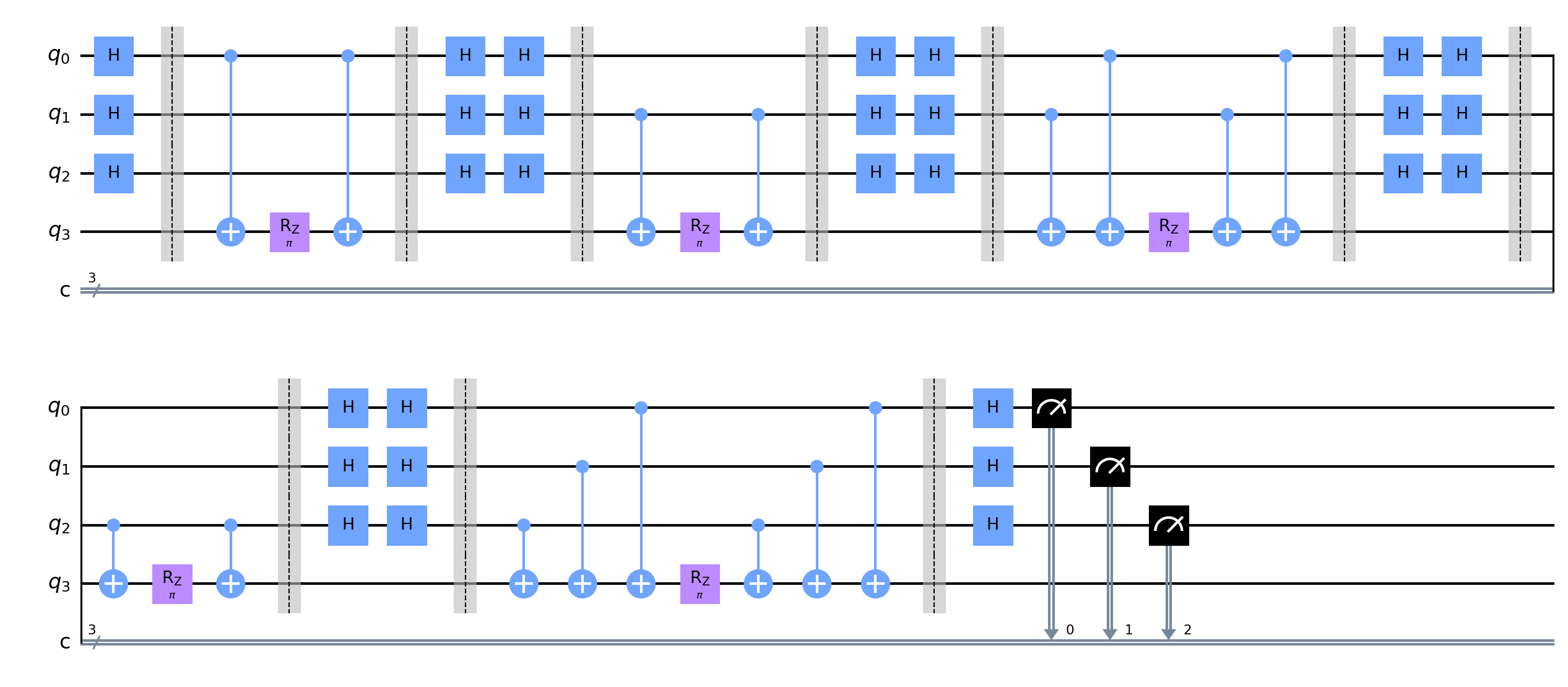}}
    \caption{Quantum circuit for $Cay(\mathbb{Z}_2^3,\{001,010,011,100,111\})$.}
    \label{fig:fig8}
\end{figure}

Recall that, if $u\oplus v=\sigma$, where $\sigma$ is given by Eq.~\ref{eq:sigma} in Theorem~\ref{thm:pst}, then $\{u,v\}$ is the PST pair. This partitions the vertex set into PST pairs. The graph shown in Fig.~\ref{fig:Q3} admits PST between pairs $\{000,111\}$, $\{001,110\}$, $\{010,101\}$, $\{011,100\}$, and the other graph in Fig.~\ref{fig:AQ3} has PST pairs $\{000,011\}$, $\{001,010\}$, $\{100,111\}$, $\{101,110\}$. Since weighted cubelike graphs, as described in Theorem~\ref{thm:pst}, are vertex-transitive, the study of PST between the pair $\{\textbf{0},\sigma\}$ is equivalent to any other pair. Therefore, every quantum circuit is initialized to state $\ket{0}^{\otimes n}$, see Fig.~\ref{fig:fig7} and Fig.~\ref{fig:fig8} which illustrate quantum circuits for the above graphs mentioned. 

\begin{figure}
    \centering
    \scalebox{.6}{\includegraphics[width=20cm]{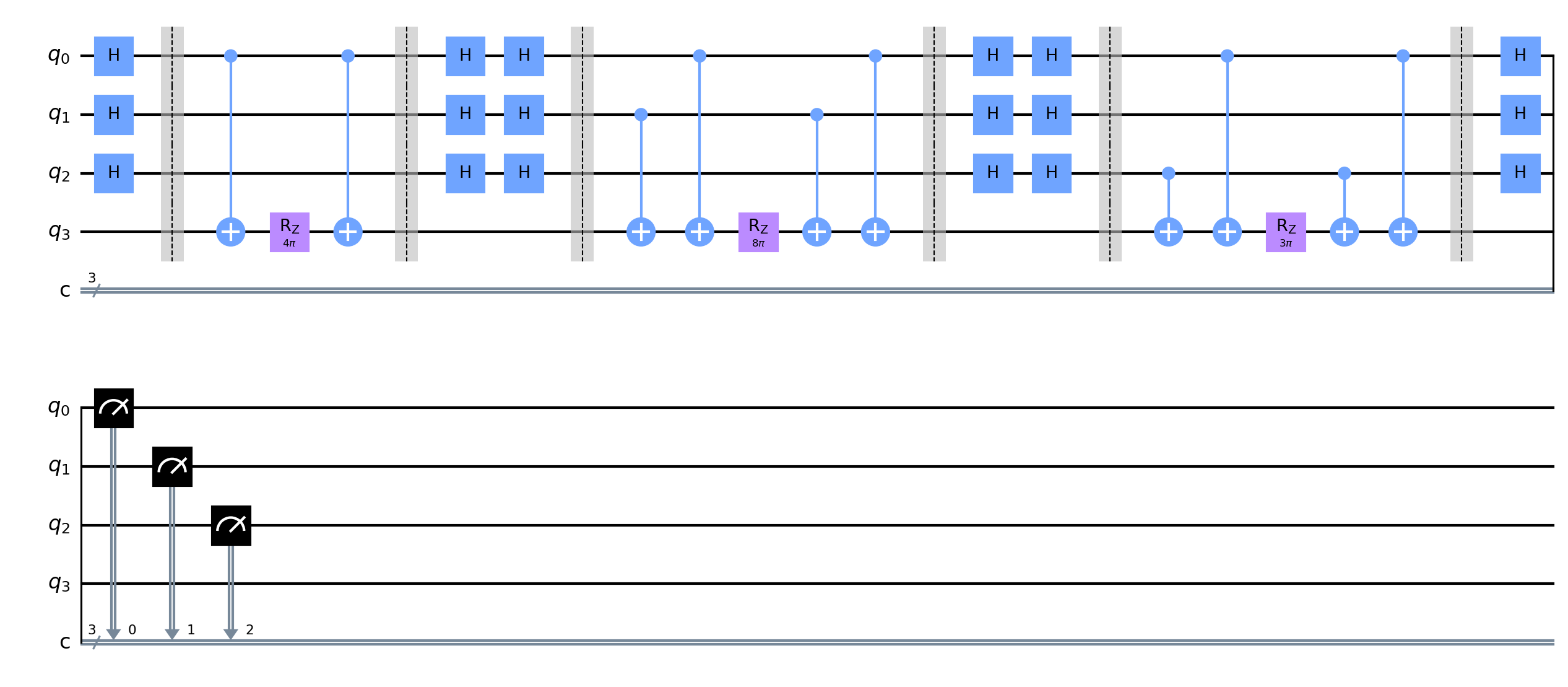}}
    \caption{Quantum circuit for $Cay(\mathbb{Z}_2^3,\{f(001)=4,f(011)=8,f(101)=3\})$.}
    \label{fig:fig9}
\end{figure}

Suppose the weight function $f$ is defined by
\begin{equation}
f(001)=4,\; f(011)=8, \; \mbox{and },\; f(101)=3,
\end{equation}
and zero on other elements, then the $3$-tuple $\sigma$ is computed as (using Theorem~\ref{thm:pst});
\begin{equation*}
    \begin{split}
    O_{001}=\{001,011,101\} &\implies f(001)+f(011)+f(101)\mbox{ mod }2=1 \\ &\implies \sigma_1=1 \\
    O_{010}=\{011\} &\implies f(011) \mbox{ mod }2=0 \\ &\implies \sigma_2=0 \\
    O_{100} = \{101\} & \implies f(101) \mbox{ mod }2=1 \\ &\implies \sigma_3=1
    \end{split}
\end{equation*} 
Thus, $\sigma=101$ and $\{000,101\}$ is a PST pair. The same is obtained by simulating the quantum circuit shown in Fig.~\ref{fig:fig9}.
On the other hand, if $f$ is defined by
\begin{equation}\label{eq:AQ3_f}
f(010)=4,\; f(011)=7,\; f(100)=8, \; f(101)=2, \; f(110)=5,
\end{equation}
then $\sigma=101$, and $\{000,101\}$ is a PST pair.

\begin{remark}
    Given a pair in a cubelike graph, we can assign weights to edges such that PST occurs between the given pair.
\end{remark}

\begin{figure}
    \centering
    \begin{minipage}{.4\textwidth}
    \centering
        \includegraphics[scale=.25]{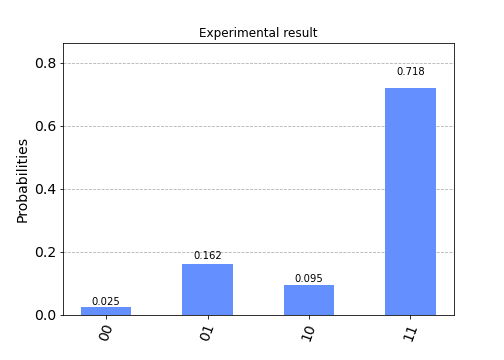}
    \end{minipage}
    \begin{minipage}{.4\textwidth}
    \centering
        \includegraphics[scale=.25]{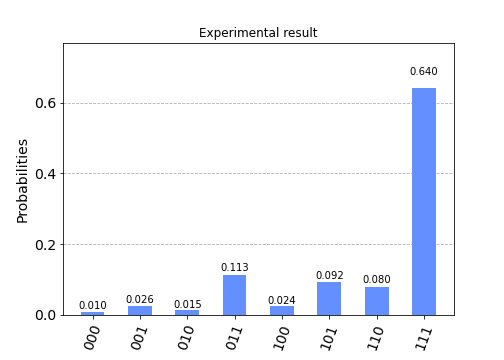}
    \end{minipage}
\caption{Experimented probability distribution of CTQW on $Cay(\mathbb{Z}_2^3,\{01,10\})$ (left) and on $Cay(\mathbb{Z}_2^3,\{001,010,100\})$ (right) after time $\frac{\pi}{2}$.}
\label{fig:fig10}
\end{figure}

\begin{note}
Quantum circuits displayed in Fig.~\ref{fig:gen_circuit} can not be run on real quantum computers due to some techincal issues such as quantum decoherence and state fidelity. We have, however, tested small graphs on the computer \textit{ibmq\_manila} as shown in Fig.~\ref{fig:fig10}.
\end{note}

\section{Conclusion and future work}
In this paper, we have experimentally tested perfect state transfer on IBM's quantum simulators and quantum computers on weighted cubelike graphs. We have used Hamiltonian simulation techniques to construct efficient circuits for continuous-time quantum random walks. We have verified the theoretical results of \cite{Cao2021} and \cite{rishi2021(2)} that PST or periodicity on integral weighted cubelike graphs occurs at time $t =\frac{\pi}{2}$, where weights are determined by Theorem~\ref{thm:pst}. In the future, we plan to construct efficient quantum circuits for quantum walks on weighted abelian Cayley graphs.

\bibliographystyle{acm}
\bibliography{ContinuousQuantumWalkSimulationOnCubelikeGraphs}
\end{document}